\documentclass[fleqn,twoside]{article}
\usepackage{espcrc2}
\usepackage{graphicx}

\title{\bf Properties of the $a_0$ resonances}

\author{{\bf Agnieszka Furman and 
           Leonard Le\'sniak\thanks{Talk given by L. Le\'sniak at the 
           QCD 2002 Conference, Montpellier, France, July 2--9 2002}}\\
         \addressmark{Henryk Niewodnicza\'nski Institute of Nuclear 
                      Physics, \\ PL 31-342 Krak\'ow, Poland}}

\begin{document}

\begin{abstract}

We present results following from the coupled channel model of two 
$a_0$ resonances decaying into the $\pi\eta$ and $\rm K\overline{K}$ 
mesons. 
The $a_0(980)$ resonance can be described by {\em two} distinct 
{\em poles}.
It is shown that the discrepancy in the $a_0(980)$ mass position 
between the Crystal Barrel Collaboration and the E852 Group can be 
explained and removed.
In our model with parameters fixed by the present experimental data the 
$a_0(980)$ {\em cannot} be interpreted as a bound $\rm K\overline{K}$ 
state although the $\rm K\overline{K}$ forces in the S-wave isovector 
channel are attractive.
\end{abstract}

\maketitle

\section{Introduction}

Among the scalar mesons there are only two isovector resonances 
$a_0(980)$ and $a_0(1450)$ \cite{pdg02}. 
Physical properties of both mesons are, however, not well known, for 
example the widths and branching ratios are poorly determined. 
Also the mass determination is problematical. 
The proper interpretation of the resonant states can be obtained in 
terms of the S-matrix poles but the data are rarely analysed in such 
a way. 
Two $\rm K\overline{K}$ thresholds at $m_{\rm K^+}+m_{\rm K^-}=987.4$ 
MeV and at $2~m_{\rm K^0}= 995.3$ MeV lie close to $a_0(980)$. 
Therefore a description of the $a_0(980)$ line shape using the simple 
Breit-Wigner formula leads to a {\em distortion} of the $a_0(980)$ mass 
and width (compare, for example differences between the results given 
in \cite{CB98} and \cite{BNL99}).
In \cite{montanet00} Montanet indicated important differences of the 
$a_0(1450)$ masses and widths found in the $p\overline{p}$ annihilation 
by the Crystal Barrel Collaboration \cite{CB98} and by the OBELIX 
Collaboration \cite{OB98}. 
Recently the WA102 Collaboration has observed the $a_0(980)$ production 
in the pp central collisions at 450 GeV, however the $a_0(1450)$ 
resonance has not been seen \cite{WA102}.

The main decay channels of the $a_0$ resonances are $\pi\eta$ and 
$\rm K\overline{K}$. 
In the first part of this paper the $a_0(980)$ production and decays 
will be discussed using the Flatt\'e model \cite{flatte76}. 
Next we shall present the results following from the coupled channel 
model formulated in 1996 \cite{lesniak96} and further developed 
in \cite{af}.

\section{How the $a_0(980)$ splits into two poles?}

Two experimental collaborations have used the Flatt\'e model to analyse
their data [2, 3].
Two different $\rm K\overline{K}$ thresholds have not been 
distinguished. 
In the Flatt\'e model the effective mass distribution in the 
$j-$channel ($j=\pi\eta,\,\rm K\overline{K}$) is given by 
 
\begin{equation}
\frac{d\sigma_j}{dm}=c |\;\!F_j\;\!|^2,
\end{equation}
where c is a constant and the production amplitudes are given by
\begin{equation}
F_j=\frac{\sqrt{\Gamma_j}}{{m_R}^2- m^2-i\,m_R\,(\Gamma_{\pi\eta}+
\Gamma_{\rm K\overline{K}})}.
\end{equation}
The $\pi\eta$ partial width is a product of the coupling constant
$g_{\pi\eta}$ and the $\pi\eta$ momentum $k_1$: 
$\Gamma_{\pi\eta}=g_{\pi\eta}\,k_1$. 
Similarly above the $\rm K\overline{K}$ threshold 
$\Gamma_{\rm K\overline{K}}= g_{\rm K\overline{K}}\,k_2$.
However, below the threshold this width becomes imaginary: 
$\Gamma_{\rm K\overline{K}}=i\,g_{\rm K\overline{K}}\,|k_2\;\!|$, where
$|k_2\;\!|=\sqrt{{m_K}^2-m^2/4}$ and $m_K$ denotes the kaon mass.
Near the $\rm K\overline{K}$ threshold $\Gamma_{\pi\eta}$ varies slowly
and $\Gamma_{\rm K\overline{K}}$ varies {\em rapidly}. 
Therefore the $\pi\eta$ distribution is {\em narrowed} on both sides of 
the $\rm K\overline{K}$ threshold as compared to the Breit-Wigner shape 
characteristic for the constant width (independent on $m$). 
The second important feature of the Flatt\'e model is the existence of 
{\em two} complex poles of the production amplitudes $F_i$ at different 
energies corresponding to the same meson $a_0(980)$.

Let us derive the pole positions following from the parameter values 
used by the E852 Group in their analysis of the reaction
$ \pi^-p\to \eta\pi^+\pi^- n$ \cite{BNL99}. 
At the beginning one has to correct the coupling constants for the 
finite experimental energy resolution. 
Then the $\pi\eta$ coupling constant reduces to the value 
$g_{\pi\eta}=0.210 \pm 0.015$ from which we have found two poles
at $E_1=(1006 -i\,25)$ MeV on sheet II and $E_2=(988 -i\,44)$ MeV on 
sheet III. 
These values are in very good agreement with the pole positions of the 
Crystal Barrel Collaboration recalculated and {\em corrected} by us: 
$E_1=(1005 -i\,25)$ MeV on sheet II and $E_2=(985 -i\,46)$ MeV on sheet 
III.
The energy differences are indeed quite large: 
$Re E_1-Re E_2= 20$ MeV; the corresponding difference of 
the total widths is $\Gamma_2-\Gamma_1= 43$ MeV. 
\begin{figure}[thb]
\vspace{-1cm}
\includegraphics*[width=18pc]{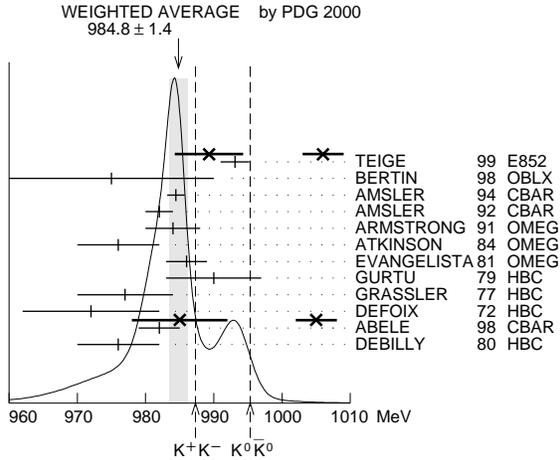}
\vspace{-1cm}
\caption{$a_0(980)$ mass determination from \cite{PDG00} supplemented
 by our determination, indicated by crosses and described in the text}
\label{fig:pdg}
\end{figure}
In Fig. 1 we show crosses with errors corresponding to our two pole 
determination of the real energy parts for the experiments \cite{CB98} 
and \cite{BNL99}. 
One can notice that about 3$\sigma$ discrepancy between the E852 value, 
based on the Breit-Wigner form, and the mean value determination by the 
Particle Data Group \cite{PDG00} disappears when one considers the pole 
value on sheet II at 988 MeV.
 
By looking at Fig. \ref{fig:pdg} one can try to answer often asked 
question: where is located the $a_0(980)$, below or above the 
$\rm K\overline{K}$ threshold? 
The most probable answer is: the first $a_0(980)$ pole on sheet III is
located below and the second $a_0(980)$ pole on sheet II is located 
above the $\rm K^0 \, \overline{K} \, \! ^0$ threshold. 
One remark is relevant here: it would be useful to print in the Review 
of Particle Physics the $a_0(980)$ masses and its widths in the form of 
the complex energy values corresponding to different determinations of 
the T-matrix poles.
Such a presentation is already given for the $f_0(600)$ and the
$f_0(1370)$ [1, 10]. 

The $a_0(980)$ coupling to the $\rm K\overline{K}$ channel is 
responsible for the appearance of two $a_0(980)$ poles. 
If $g_{\rm K\overline{K}}=0$ then $E_2=E_1$. 
If, however, $g_{\rm K\overline{K}}\neq 0$ then
\begin{equation}
Re E_2-Re E_1 \approx -g_{\rm K\overline{K}}\;g_{\pi\eta}\,\frac{m_K\,q_1}{4\,Re q_2}
\end{equation}
and
\begin{equation}
\Gamma_2-\Gamma_1 \approx 2\,g_{\rm K\overline{K}}\,Re q_2. 
\end{equation}
In these equations $q_1$ is the $\pi\eta$ relative momentum at the 
$\rm K\overline{K}$ threshold and $q_2$ denotes the $\rm K\overline{K}$
momentum at the $a_0(980)$ pole on sheet III.  
\section{Coupled channel model of the $a_0(980)$ and the $a_0(1450)$
resonances}

Below we present results of the simple two-channel model of $a_0$ 
resonances applied in \cite{af} to the analysis of experimental results. 
In this model four reactions: $\pi\eta \to \pi\eta$,
$\pi\eta \to \rm K\overline{K}$, $\rm K\overline{K}\to \pi\eta$ and
$\rm K\overline{K}\to \rm K\overline{K}$ are described simultaneously
using the separable interactions in the form     
\begin{equation}
\langle\,\mathbf{p}\,|\,V_{ij}\,|\,\mathbf{q}\,\rangle=
\lambda_{ij}\,f_i(p)\,f_j(q)\;, \qquad i,j=1,2\;.
\end{equation}
Here $\lambda_{ij}$ are the real coupling constants and $f_i$ are the 
Yamaguchi form factors inversely proportional to $p^2+\beta_i^2$, where 
$p$ is the c.m. momentum and  $\beta_i$ are constants. 
The model has altogether only five independent parameters: the 
$\pi\eta$ coupling constant $\lambda_{11}$, the $\rm K\overline{K}$ 
coupling constant $\lambda_{22}$, the interchannel coupling 
$\lambda_{12}$ and two range parameters $\beta_1$ and $\beta_2$. 
The T-matrix satisfies the Lippmann--Schwinger equation $T=V+V\,G\,T$, 
where G is the propagator matrix.

We fix four model parameters by choosing the $a_0(980)$ pole of the
T-matrix on sheet II at $(1005-i\,24.5)$ MeV and the $a_0(1450)$ pole
on sheet III at $(1474-i\,132.5)$ MeV. 
The fifth model parameter is constrained by the experimental 
$\rm K\overline{K}$/$\pi\eta$ branching ratio near the 
$\rm K\overline{K}$ threshold
\begin{equation}
 BR=\frac{\int_{m_2}^{m_{max}} \rho_2 \,|F_2(m)|^2 \,dm }
   {\int_{m_1}^{m_{max}} \rho_1 \,|F_1(m)|^2 \,dm },
\end{equation}
where $\rho_i=2k_i/m$. 
We have chosen $m_1=m_{\pi^0}+m_{\eta}$ and $m_2=2\,m_{\rm K^0}$, 
corresponding to the branching ratio for the neutral $a_0(980)$ decays. 
The branching ratio depends very strongly on the upper integration 
limit $m_{max}$ as shown in Fig. 2. 
The Crystal Barrel result \cite{CB98} $BR=0.23\pm0.05$ corresponds to 
large value of $m_{max}=2\,m_p-m_{\pi^0}=1741$ MeV. 
The upper limit of the WA102 group \cite{WA102} is lower as it is equal 
to $1147$ MeV since this group studied the $f_1(1285)$ decay into 
$\pi\pi\eta$. 
Their measured value $0.166\pm 0.01\pm 0.02$ agrees well with the 
theoretical value 0.19 shown in Fig. 2. The OBELIX ratio $0.26\pm0.06$
\cite{OB98} is also in a good agreement with the theoretical curve.
\begin{figure}[htb]
\includegraphics*[width=14pc]{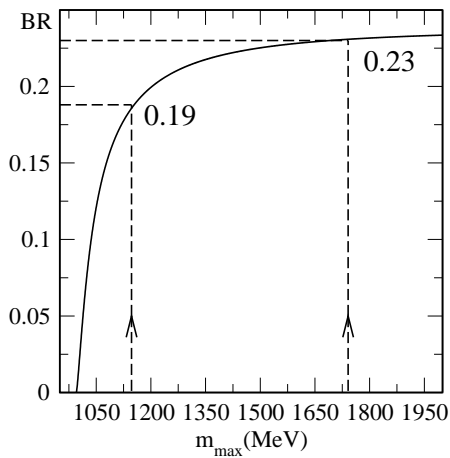}
\vspace{-.5cm}
\caption{Dependence of the $a_0(980)$ decay branching ratio on the
 effective mass upper limit}
\label{fig:rozpady}
\vspace{-.6cm}
\end{figure} 
%
\section{Model prediction in the $a_0(1450)$ range}
According to \cite{PDG00} we take the $a_0(1450)$ mass equal to 
$M=1474$ MeV and its total width $\Gamma=265$ MeV. 
Then the $\rm K\overline{K}$/$\pi\eta$ branching ratio calculated in 
the limits of $m$ between $M-\Gamma/2$ and $M+\Gamma/2$ is equal to 
0.98. 
When this ratio is evaluated in the slightly larger limits between 
$1300$ MeV and $1741$ MeV then it decreases to $0.78$. 
Both numbers stay well within the experimental value $0.88\pm 0.23$ 
found by the Crystal Barrel Collaboration \cite{CB98}. 
Thus the model provides us with the theoretical branching ratios 
consistent with the  experimental findings for both $a_0$ mesons.
%
\section{Other results following from the model}
There are many other predictions coming from the coupled channel model
described previously. 
One can calculate the elastic and transition amplitudes, the effective 
mass distributions in two channels, positions of other T-matrix poles 
not initially imposed in the procedure of fixing the model parameters 
and the coupling constants in two channels at each pole. 
Some of these predictions have been already published in \cite{af}. 
We have calculated rather important energy difference of the $a_0(980)$ 
poles: $Re E_1-ReE_2=13.5$ MeV and the width difference 
$\Gamma_2-\Gamma_1=18$ MeV. 
These numbers are in qualitative agreement with the results found in 
Sect. 2 using the Flatt\'e model. 

\begin{figure}[htb]
\vspace{-.6cm}
\includegraphics*[width=15pc]{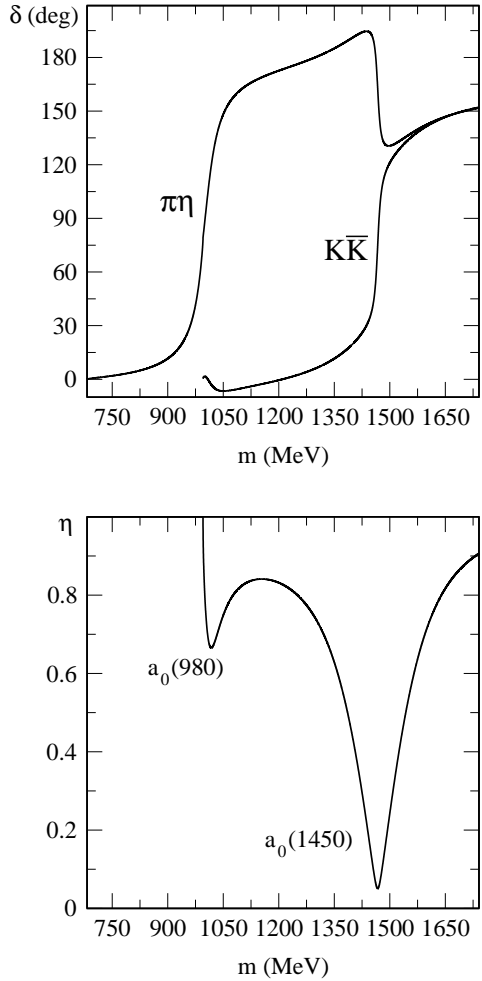}
\vspace{-.5cm}
\caption{Channel phase shifts and inelasticity versus the effective 
mass}
\label{fig:phase}
\vspace{-1cm}
\end{figure}

Three interesting physical quantities are related to the diagonal
S-matrix elements in two channels:
\begin{equation}
S_{jj}= \eta e^{2\,i\delta_j},\qquad j=1,2. 
\end{equation}
The inelasticity $\eta$ is common for both channels, what follows from
the unitarity condition, however the phase shifts $\delta_i$ are
different in both channels, as shown in Fig. \ref{fig:phase}. 
The sudden rise of the $\pi\eta$ phase shifts near the 
$\rm K\overline{K}$ threshold is due to the presence of the $a_0(980)$ 
pole on sheet II. 
Rather weak variation of the $\rm K\overline{K}$ phase shifts near $1$ 
GeV is a result of the destructive interference between two related 
$a_0(980)$ poles lying on sheets II and III. 
The interference effects are also seen near $1450$ MeV where in the 
$\pi\eta$ channel the $a_0(1450)$ pole interferes with the $S_{11}$ 
zero related to the second $a_0(1450)$ pole. 
In the $\rm K\overline{K}$ channel, however, the interference is 
constructive and the $\rm K\overline{K}$ phase shifts rise very quickly
in vicinity of $1500$ 
MeV. The presence of the $a_0$ resonances leads to two characteristic 
dips of the inelasticity coefficient $\eta$.   
%
\section{Nonexistence of the bound $\rm K\overline{K}$ isovector state}

The present model is suitable to study the question whether a 
$\rm K\overline{K}$ pair can form a bound S-wave isospin $1$ state. 
This question is relevant since the coupling between kaons in that 
state is negative so the forces in the isovector state are attractive. 
In order to answer this interesting question we have studied the 
evolution of the T-matrix poles in the limit of vanishing interchannel 
coupling constant $\lambda_{12}$. 
In this limit one can find that the $\rm K\overline{K}$ resonance 
exists at the complex effective mass equal to $(1270-i\,77)$ MeV. 
If the interchannel coupling is switched on then this resonance evolves 
into the $a_0(1450)$ pole and not into the $a_0(980)$ pole. 
The energy of the bound state would be real and smaller than the sum of 
two kaon masses. 
Since this is not a case in our model the answer about the
presence of the bound $\rm K\overline{K}$ isovector state is 
{\em negative}.  
%
\section{Conclusions}

We have found that the $a_0(980)$ meson can be described in terms of 
two distinct poles lying near the $\rm K\overline{K}$ threshold on 
sheets II and III. 
This result is a common feature of the two models, namely the 
Flatt\'e model, often used to describe a single $a_0$ resonance, and 
the coupled channel model of two $a_0$ resonances, constructed recently 
by us. 
The second model has been constrained in the $a_0(980)$ mass range and 
then successfully applied at higher effective $\pi\eta$ or 
$\rm K\overline{K}$ masses where the $a_0(1450)$ resonance is 
present.
The two channel model can be extended to treat more decay channels.


\end{document}